\documentclass[useAMS,usenatbib,usegraphicx]{mn2e}
\newcommand{\beq}[1]{\begin{equation}\label{#1}}
 \newcommand{\eeq}{\end{equation}}
 \newcommand{\bea}{\begin{eqnarray}}
 \newcommand{\eea}{\end{eqnarray}}
\def\la{\mathrel{\hbox{\rlap{\hbox{\lower4pt\hbox{$\sim$}}}\hbox{$<$}}}}

\usepackage{epsf}
\usepackage{amssymb,amsmath}

\title[Consistent Test of the DD Relation]{A Consistent Test of the Distance-Duality Relation with
Galaxy Clusters and Type Ia Supernave }

\author[Liang et al.]
{Nan Liang$^{1,2,3}$\thanks{liangn@bnu.edu.cn}, Zhengxiang Li$^2$,
Puxun Wu$^4$, Shuo Cao$^2$,  Kai Liao$^2$,  and Zong-Hong
Zhu$^2$\thanks{zhuzh@bnu.edu.cn}
\\$^1$Center for High Energy Physics, Peking University, Beijing 100871,  China
\\$^2$Department of Astronomy, Beijing Normal University, 100875,
Beijing, China
\\$^3$National Astronomical Observatories, Yunnan Observatory, Chinese
Academy of Sciences, Kunming 650011, China
\\$^4$Center for Nonlinear Science and Department of Physics, Ningbo
University, Ningbo, Zhejiang 315211, China
}

\begin{document}

\date{}

\voffset- .5in

\pagerange{\pageref{firstpage}--\pageref{lastpage}} \pubyear{}

\maketitle

\label{firstpage}

\begin{abstract}
We propose a  new consistent method to test the distance-duality
(DD) relation which related the angular diameter distances ($D_{A}$)
to the luminosity distances ($D_{L}$) in a cosmology-independent
way. In order to avoid any bias brought by redshift incoincidence
between galaxy clusters and  Type Ia Supernave (SNe Ia), as well as
to ensure the integrity of the galaxy clusters samples, we obtain
the luminosity distance of a certain SN Ia point at the same
redshift of the corresponding galaxy cluster by interpolating from
the nearby SNe Ia. With the observational data pairs at the same
redshifts of the angular diameter distances from the complete 38
galaxy clusters for the spherical model and the corresponding
luminosity distances interpolated from the Union2 set, we find that
$\eta \equiv {D_{L}}{(1+z)}^{-2}/{D_{A}}=1$ is satisfied within
$2\sigma$ confidence level for parameterizations of $\eta(z)$, which
are more stringent than previous results without considering
redshift bias.
\end{abstract}


\section{Introduction}
\label{sec:introduction}

The Distance-duality (DD) relation is known as the Etherington's
reciprocity relation \citep{eth33}, which is related the angular
diameter distance (ADD, $D_{A}$) to the luminosity distance
($D_{L}$)  by means of a single parameter,
\begin{equation}
  \eta \equiv \frac{D_{L}}{D_{A}}{(1+z)}^{-2}=1.
  \label{rec1}
\end{equation}
This equation  is completely valid for all cosmological models based
on Riemannian geometry \citep{Ellis07}. Therefore, the DD relation
plays an essential role in observational astrophysics and modern
cosmology, 
such as galaxy clusters observations \citep{Lima03,Cunha07},  the
anisotropies of cosmic microwave background (CMB) \citep{Komatsu11},
as well as gravitational lensing studies \citep{Schneider99,Fu08}.
In principle, if both $D_{A}$  and $D_{L}$ of cosmological sources
at the common redshifts are known, the DD relation ($\eta=1$) could
be directly tested by means of astronomical observations.  From
Sunyaev-Zeldovich effect (SZE)\citep{Sunyaev} and X-ray surface
brightness of galaxy clusters, the observational ADDs of galaxy
clusters  can be obtained \citep{Silk78}. By using an isothermal
spherical model for which the hydrostatic equilibrium model and
spherical symmetry assumed, \cite{Reese02} selected 18 galaxy
cluster sample and \cite{Mason01}  obtained seven clusters from
the X-ray-limited flux sample. 
The measurements of the two samples above have been corrected  by
using an isothermal elliptical model to get 25 ADDs of galaxy
clusters \citep{DeFilippis05}. Recently,  \cite{Boname06} obtained
38 galaxy clusters sample by assuming the spherical model. 

\cite{Uzan04} considered ADDs of 18 galaxy cluster sample
\citep{Reese02}  to test the DD relation by assuming the
$\Lambda$CDM model via the technique,
$D_A^{\mathrm{cluster}}(z)=D_A^{\Lambda \rm CDM}(z)\eta^2(z)$, and
showed that no violation of the DD relation is only marginally
consistent. \cite{DeBernardis06}  considered 38 galaxy cluster for
spherical model \citep{Boname06} to test the DD relation by assuming
the $\Lambda$CDM model. In order to test the DD relation in a
model-independent way, one should use measurements of $D_L$ such as
Type Ia Supernave (SNe Ia) directly. By binning $D_L$ of SN Ia data
and ADDs from FRIIb radio galaxies and ultra compact radio sources,
\cite{Bassett04} found that the brightening excess of SNe Ia at $z
> 0.5$ could cause a moderate violation at 2$\sigma$ confidence
level (CL). \citet{DeBernardis06} binned ADD data of galaxy clusters
\citep{Boname06} and SNe Ia
data 
to find that the validity of $\eta = 1$ is consistent at 1$\sigma$
CL.

However, \cite{HLR10} argued that the  above tests may have been
influenced by the particular choice of redshift bin, and they tested
the DD relation with two ADD samples \citep{Boname06,DeFilippis05}
and the Constitution set of SNe Ia data \citep{Hicken09}. For the
biggest redshift difference  between clusters and SNe Ia is $\Delta
z=|z_{\rm cluters}-z_{\rm SNe}|\simeq0.01$ for three clusters, a
selection criteria ($\Delta z\le0.005$) for a given pair of data set
are used to avoid the corresponding bias of redshift differences.
With the incomplete spherical model sample \citep{Boname06} in which
three ADD data have been discarded, they found a strong violation
($>3\sigma$) of the DD relation by using two parameterizations of
$\eta$ parameter [$\eta(z)=1+\eta_1z$, and $\eta(z)=1+\eta_a
z/(1+z)$]. More recently, \cite{Li11}  used the same selection
criteria for given pairs of observational data to remove more data
points of the galaxy clusters corresponding to the Constitution set
and found that the DD relation could be marginally accommodated at
$3\sigma$ CL for the spherical model if the effect
of the errors of SNe Ia considered. 
Additionally, they also examined the DD relation for two more
general parameterizations of $\eta$ parameter
to show that $\eta(z)=1$ is
compatible with 
the spherical model sample and the Union2 set \citep{Amanullah} at 
$2\sigma$ CL. Some authors have been proposed other astrophysical
sources  in context of testing the
DD relation, 
such as the baryon acoustic oscillation (BAO) observation
\citep{More09, Cardone12}, the observational Hubble parameter
 data  \citep{Avg10}, as well as the X-ray gas fraction
($f_{\rm gas}$) data \citep{Gon12, HGA12}. For recent works of DD
relation on astrophysical research, see e.g. \cite{CaoZ11,Nair11},
\cite{HLR11,Lima11,HLR12}, and \cite{H12}.

It is obvious that testing results of the DD relation may be
influenced by the particular choice of the selection criteria for a
given pair of data set. The difference of redshifts between pairs of
galaxy clusters and SNe Ia may cause obvious deviation in testing
the DD relation. In principle, the only strict criterion to form a
given pair is that galaxy clusters and SNe Ia locate at exactly the
same redshift. At other hand, the more stringent selection criteria
are used, the more data points should be removed. In order to avoid
any bias of redshift differences between SNe Ia and galaxy clusters
and ensure the integrity of observational data pairs, we can use the
nearby SNe Ia points to obtain the luminosity distance of SN Ia
point at the same redshift of the corresponding galaxy cluster; this
situation is similar with the cosmology-independent calibration of
GRB relations directly from SNe Ia
\citep{Liang2008,LZ2008,Liang2010,Liang2011}.
In this work, we test the DD relation 
with the SNe Ia points in which a sub-sample are corrected by
interpolating from the nearby points to the same
redshifts of the corresponding galaxy clusters sample  
for a given pair of data set. 
We focus on the current observational data pairs of 
galaxy cluster sample under an assumption of spherical model
and the Union2 set in the present work. When considering redshift
bias of observational data pairs between the complete 38 galaxy
clusters for the spherical model and the corresponding Union2 set,
we find that $\eta(z)=1$ is satisfied within $2\sigma$ confidence
level with current observations.


\section{Data Analysis}\label{sec:analysis}
In this work, we test the DD relation with the 38 ADD sample from
galaxy clusters for the spherical model 
and the
Union2 set which consists of 557 SNe Ia. 
It is
easy to find that differences of redshifts between the 38 galaxy
clusters to the Union2 set are more centered around $\Delta z=0$ and
the biggest value at $\Delta z=0.005$ for a given pair of data set;
this situation can provide the accuracy in the interpolating
procedure. Therefore, we can obtain the luminosity distance of SN Ia
at the same redshift of the corresponding galaxy cluster by
interpolating from the nearby SNe Ia points  with the biggest
difference of redshifts $\Delta z _{\rm max}=0.005$ for a given pair
of data set. Obviously, our method can successfully avoid the
systematic errors brought by redshift incoincidence of the
observational data pairs and ensure the integrity of observational
data pairs.

When the
linear interpolation is used, 
the interpolated  distance modulus  of a source at redshift $z$ can
be calculated by
$\mu(z)=\mu_i+[(z-z_{i})/(z_{i+1}-z_i)](\mu_{i+1}-\mu_{i})$, where
$\mu_{i}$, $\mu_{i+1}$  are the distance moduli of the SNe at nearby
redshifts $z_{i}$, $z_{i+1}$; and the uncertainty is
$\sigma_{\mu}=([(z_{i+1}-z)/(z_{i+1}-z_i)]^2\sigma_{\mu,i}^2+[(z-z_{i})/(z_{i+1}-z_i)]^2\sigma_{\mu,i+1}^2)^{1/2}.$
 It is noted that some SNe data locate at the same redshifts,
therefore we weighted the SNe data at the same redshifts each other
in the interpolating procedure,
$\bar{\mu}(z)={\sum\left(\mu_i/\sigma^2_{\mu_i}\right)}/{\sum1/(\sigma^2_{\mu_i})}$,
where $\bar{\mu}(z)$ stands for the weighted mean distance modulus
at the same redshift $z$ with its uncertainty
$\sigma_{\bar{\mu}}=(\sum1/\sigma^2_{\mu_i})^{-1/2}$.

In Figure 1,
we plot $D_A$ data from the galaxy cluster 
and the corresponding corrected $D_L$ data from Union2 sub-sample at
the same redshifts of galaxy clusters. For galaxy cluster samples,
the typical statistical and systematic uncertainties of galaxy
clusters are around $\pm20\%$ and $\pm12.4\%$
\citep{Boname06,HLR10}.  Following \citet{HLR10, Li11}, we combine
the statistical and systematic uncertainties of galaxy clusters in
quadrature and treat the asymmetry uncertainties of galaxy clusters
by an statistical approach \citep{DA04}.

The technique for determining the ADDs with the SZE+X-ray
observations of galaxy clusters \citep{Sunyaev, Cavaliere} is
strongly dependent on the valid of the DD relation. 
It gives $D_A^{\mathrm{cluster}}(z)=D_A(z)\eta^2$ when the DD
relation does not hold. Therefore $D_A(z)$ must be replaced with
$D_A^{\mathrm{cluster}}(z)\eta^{-2}$ when testing the DD relation
\citep{HLR10}. The observed $\eta_{obs}(z)$ in a redshift-dependent
form can be determined by
\begin{eqnarray}
\eta_{obs}(z)=(1+z)^2\frac{{D_A^\mathrm{cluster}}(z)}{D_L^{\rm
corrected}(z)}.
\end{eqnarray}
where  $D_{A}^{\rm cluster}$ is ADD from galaxy cluster at redshift
$z$ inside the samples, and $D_{L}^{\rm corrected}$ is the corrected
luminosity distance interpolated from the nearby SNe Ia points
$D_{L}^{\rm SNe}$. We note that the data points of  SNe Ia are given
in terms of the distance modulus, which could reduce to the
luminosity distance by $ D_L(z)=10^{\mu(z)/5-5}$. Accordingly, the
uncertainty of the luminosity distance could be expressed as
$\sigma_{D_L}=(\ln10/5) D_L\cdot\sigma_{\mu}$.

\begin{figure} 
\begin{center}
\includegraphics[width=1 \hsize]{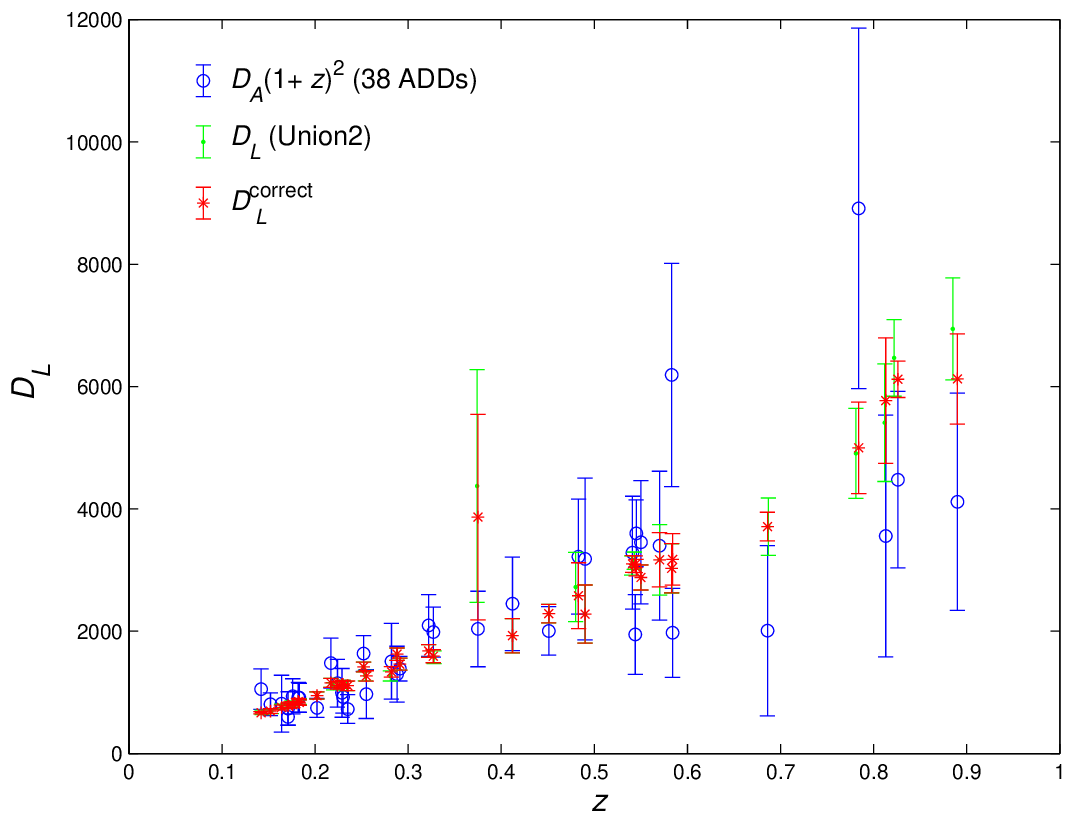}
\end {center}
 \caption{ Hubble diagram of the 38 Galaxy cluster samples for the spherical model and the corresponding SNe Ia data (Union2) with the associated error bars.
 The blue open circles   represent  $(1+z^2)D_A$ from the ADDs,
the green filled circles  represent  the corresponding  $D_L$ from
the Union2 sub-sample directly, and the red asterisks stand the
corrected $D_L$ interpolated from the nearby Union2 sub-sample.
\label{fig:hubble}}
\end{figure}

The DD relation can be tested with the combined observational data
pairs of galaxy clusters and SNe Ia  at the same redshift  by the
minimum $\chi^2$ method and the total $\chi^2$ can be given by
\begin{equation}
\chi^2(\mathbf{p})=\sum_z\frac{[\eta(z;\mathbf{p})-\eta_{\rm
obs}(z)(1+\alpha)]^2}{\sigma^2_{\eta_{\rm obs}}},
\end{equation}
where $\eta(\mathbf{p})$  represents the theoretical value with the
parameter set $\mathbf{p}$, and $\eta_{\rm obs}$ associated with the
observational technique with its error $\sigma_{\eta_{\rm
obs}}=\eta_{\rm
obs}^2[(\sigma_{D_A}/D_A^{\mathrm{cluster}})^2+(\sigma_{D_L}/D_L^{\rm
corrected})^2]$, 
the additional term $(1+\alpha)$ is introduced to take into account
a systematic uncertainty of the SN distance modulus.\footnote{We
thank the anonymous referee for pointing this out. For the absolute
magnitude of SNe is known up to $\pm0.05$ mag, the luminosity
distance can be shifted by a factor $0.023\%$.  We marginalize over
this term as a nuisance parameter with a Gaussian prior centered on
$<\alpha>=0$ and with standard deviation $\sigma_\alpha= 0.023$
\citep{Cardone12}. }

It should be noted that we do not know explicitly the difference of
redshifts between pairs of galaxy clusters and SNe Ia cause
deviation in testing the DD relation by how much, and whether the
redshift interpolation method indeed removes the bias. Therefore it
is important to check the interpolation scheme by using simulated
data  in testing the DD relation.  We list the testing results by
using the simulated data in the appendix A.\footnote{We thank the
anonymous referee for pointing this out. By using simulated data, we
find that the difference of redshifts between pairs of galaxy
clusters and SNe Ia could cause deviation in testing the DD relation
explicitly and the redshift interpolation method indeed removes the
bias of the difference of redshifts between data pairs.}

\section{Results}\label{sec:Results}

\begin{figure*}
\begin{center}
\includegraphics[width=0.35\hsize]{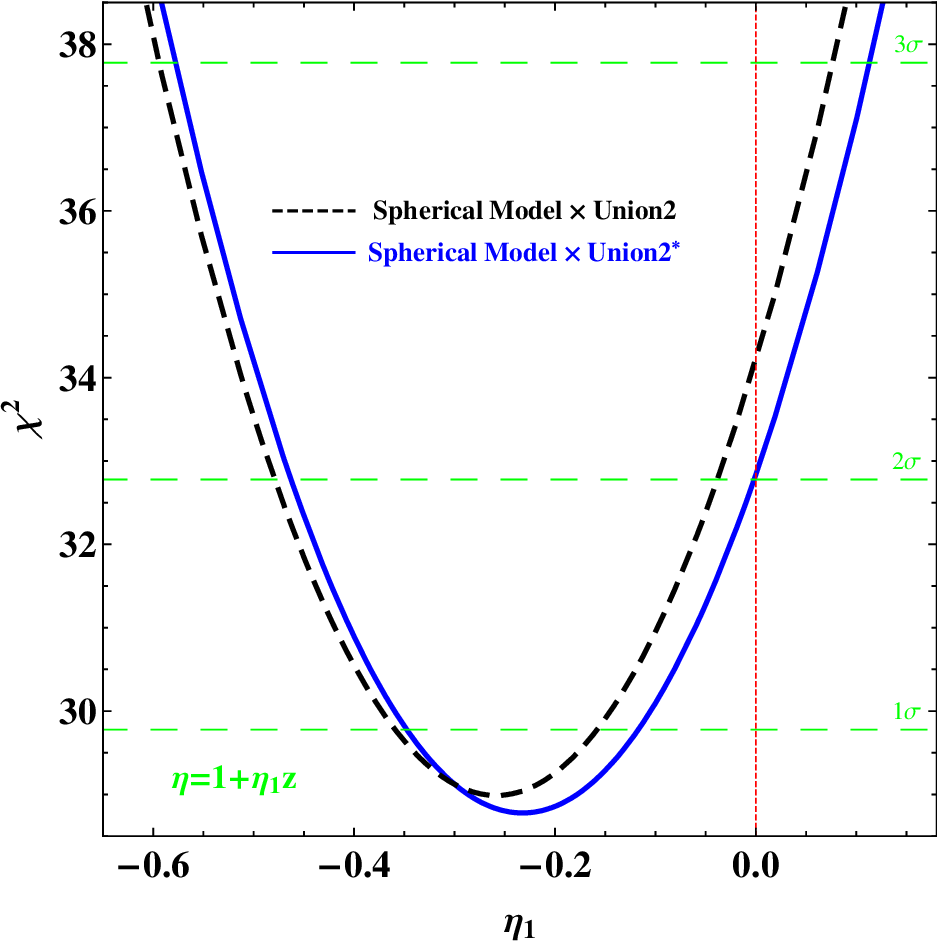}
\includegraphics[width=0.35\hsize]{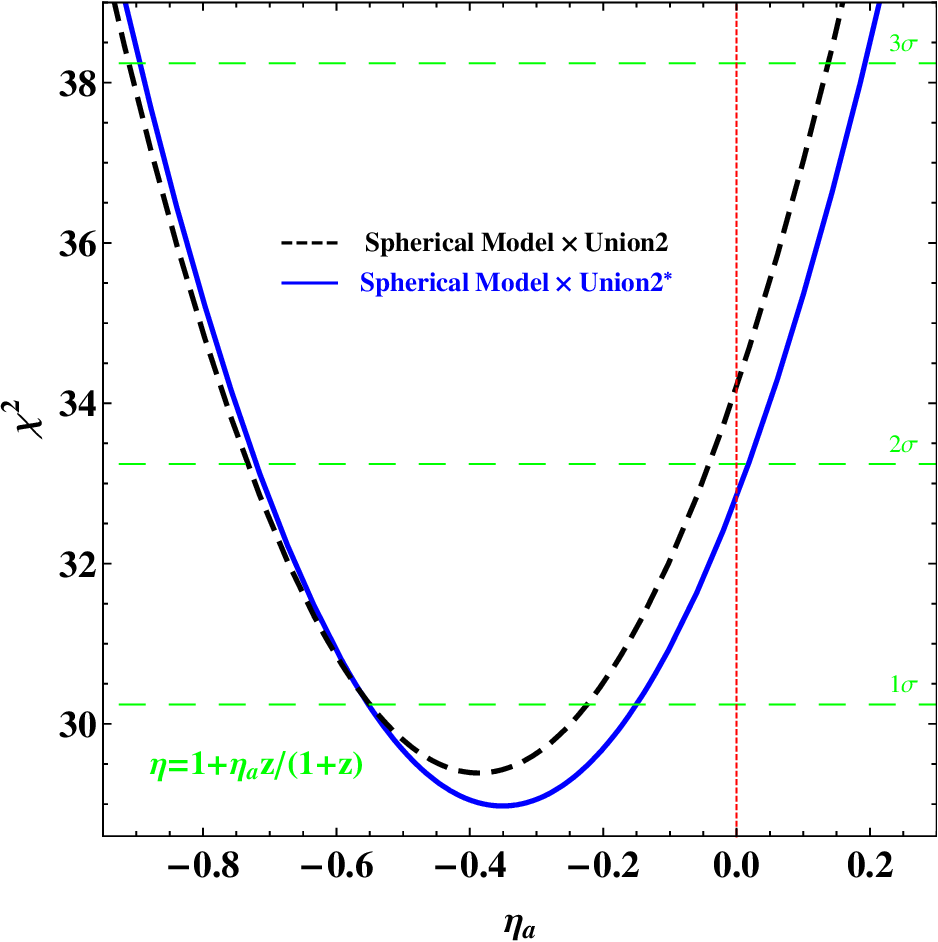}
\end {center}
 \caption{ Likelihood contours  with the 38
ADDs of galaxy clusters for spherical model and the corrected
luminosity distances of the Union2 set in the $\eta_1-\Delta\chi^2$
plane (\textit{Left}: for $\eta(z)=1+\eta_a z$), and in the
$\eta_a-\Delta\chi^2$ plane (\textit{Right}: for
$\eta(z)=1+\eta_a\frac{z}{1+z}$ ).
 The blue real lines represent the case with the corrected
luminosity distance interpolated from the nearby SNe Ia (Union2*),
the black dashed lines represent the case with the SNe Ia set
(Union2) directly, and the red vertical lines represent $\eta(z)=1$.
\label{fig:Union2}}
\end{figure*}

We show testing results of the DD relation with ADDs and the Union2
set  by considering one-parameter
parameterizations [$\eta(z)=1+\eta_1z$ 
and $\eta(z)=1+\eta_az/(1+z)$] in Figure \ref{fig:Union2}. 
For comparison, the case with  the corrected luminosity distance
(${D_L^{\rm corrected}}$) interpolated from the nearby SNe Ia and
the case with the Union2 set directly (${D}^{\rm SNe}_L$) are given
simultaneously. 
For the case with ${D_L^{\rm corrected}}$, the best-fit values are
$\eta_1=-0.232\pm0.232$ at $2\sigma$ CL with $\chi^2_{\rm
min}=28.78$, and $\eta_a=-0.351\pm{0.368}$  ($2\sigma$) with
$\chi^2_{\rm min}=28.98$. For the case with the Union2 set directly,
the best-fit values are
$\eta_1=-0.257\pm0.225$ ($2\sigma$), 
and $\eta_a=-0.387\pm0.353$ ($2\sigma$), which  are consistent with
those obtained by \citet{Li11} ($\eta_1=-0.22\pm{0.21}$ and $\eta_a=-0.33\pm{0.33} (2\sigma)$). 
We summarize the testing results  with the 38 ADDs of galaxy
clusters and the corrected luminosity distances of the Union2 set in
Table 1.\footnote{In order to find the impact of the extra factor
$(1+\alpha)$ in Equation (3), we also obtain results without the
additional term. For comparing to results with and without the
factor, we could find that the factor does NOT affect the testing
results significantly.}

\begin{figure*}
\begin{center}
\includegraphics[width=0.35\hsize]{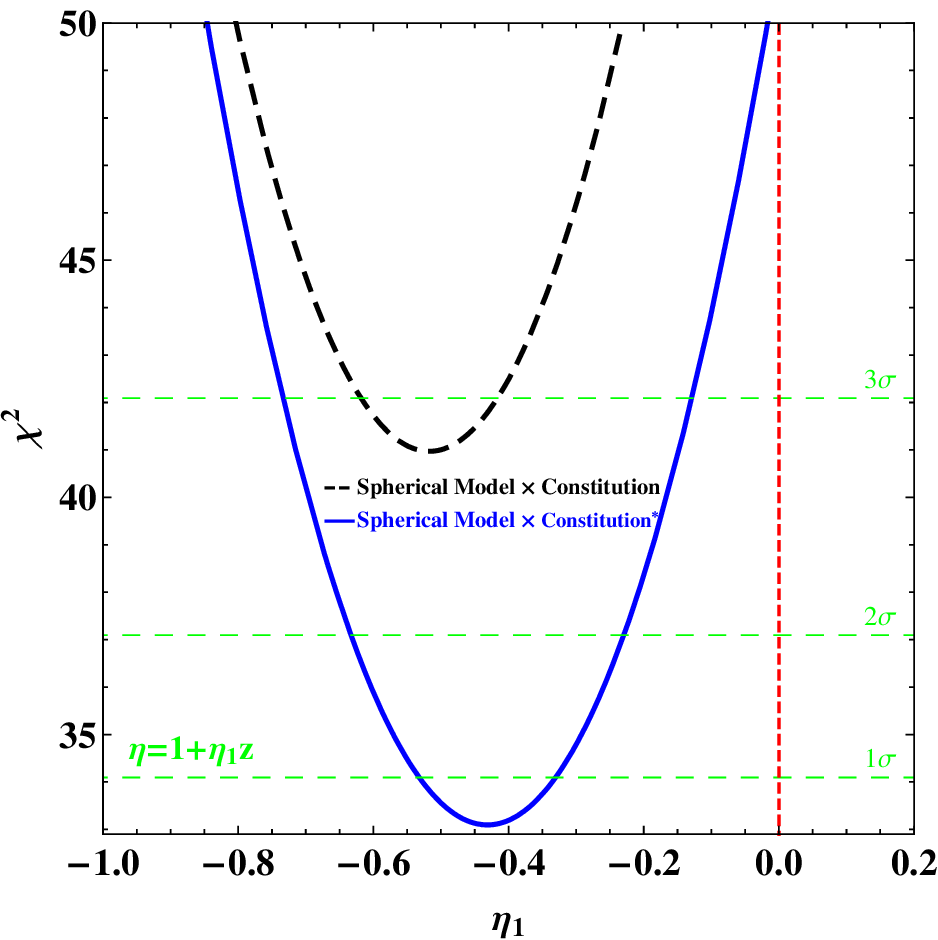}
\includegraphics[width=0.34\hsize]{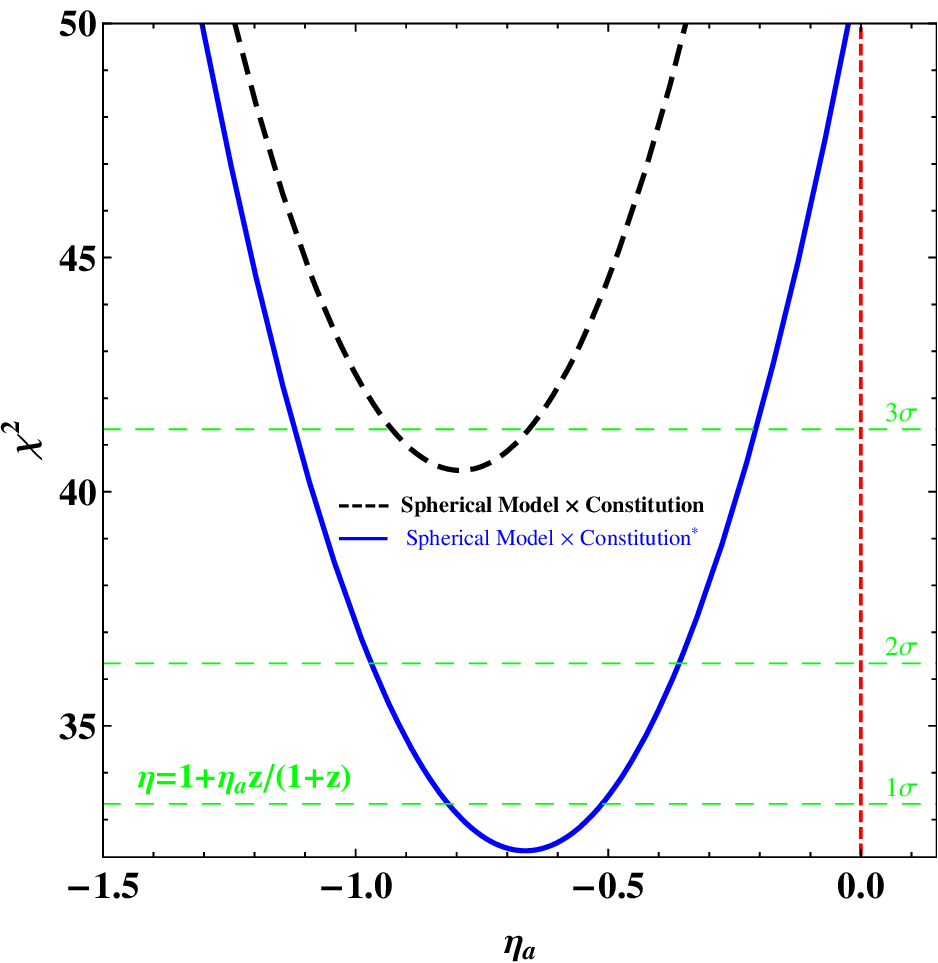}

 \end{center}
\caption{Likelihood contours  with the 38 ADDs of galaxy clusters
for spherical model and the corrected  luminosity distances of the
Constitution set  in the $\eta_1-\Delta\chi^2$ plane (\textit{Left}:
for $\eta(z)=1+\eta_1 z$),
 and in the $\eta_a-\Delta\chi^2$ plane (\textit{Right}: for $\eta(z)=1+\eta_a\frac{z}{1+z}$ ).
 The blue real lines represent the case with the corrected
luminosity distance interpolated from the nearby SNe Ia
(Constitution*),  the black dashed lines represent the case with the
SNe Ia set (Constitution) directly, and the red vertical lines
represent $\eta(z)=1$.
 \label{fig:Constitution}}
\end{figure*}

In order to compare with previous results from the incomplete ADD
sample and the Constitution set, we also show testing results with
complete ADD sample and the Constitution set in Figure
\ref{fig:Constitution}. For the case with ${D_L^{\rm corrected}}$,
the best-fit values are $\eta_1=-0.431\pm0.303$ at $3\sigma$ CL with
$\chi^2_{\rm min}=33.10$, and $\eta_1=-0.664\pm0.457$ ($3\sigma$)
with $\chi^2_{\rm min}=32.34$. For the case with the Constitution
set directly, the best-fit values are $\eta_1=-0.517\pm0.286$
($3\sigma$), 
and $\eta_a=-0.793\pm0.436$ ($3\sigma$). 
Results with the Constitution set are summarized in Table
1.\footnote{For simplicity, we do not consider the additional term
$(1+\alpha)$ in the $\chi^2$ function for the case with the
Constitution set.}

\begin{table}
{\scriptsize
\begin{tabular}{|l|c|c|c|}
\hline\hline
Parameterization (SN Ia*)  ~&$\eta_1{/}\eta_a$  ~&$\chi^2_{\rm min}$&$\chi^2_{\rm dof}$
~~\\
\hline
\hline $1$+$\eta_1z$ (Union2*) &$\eta_1$=$-0.232$$\pm{0.232}(2\sigma)$~&$28.78$~&~$0.778$~\\
\hline $1$+$\eta_1z$ (Union2)  &$\eta_1$=$-0.257$$\pm{0.225}(2\sigma)$~&$28.99$~&~$0.783$~\\
\hline $1$+$\eta_a\frac{z}{1+z}$ (Union2*) &$\eta_a$=$-0.351$$\pm{0.368}(2\sigma)$~&$28.98$~&~$0.783$~\\
\hline $1$+$\eta_a\frac{z}{1+z}$ (Union2)  &$\eta_a$=$-0.387$$\pm{0.353}(2\sigma)$~&$29.39$~&~$0.794$~\\
\hline
\hline $1$+$\eta_1z$ (Constitution*) &$\eta_1$=$-0.431$$\pm{0.303}(3\sigma)$~&$33.10$~&~$0.895$~\\
\hline $1$+$\eta_1z$ (Constitution) &$\eta_1$=$-0.517$$\pm{0.286}(3\sigma)$~&$40.97$~&~$1.107$~\\
\hline $1$+$\eta_a\frac{z}{1+z}$ (Constitution*) &$\eta_a$=$-0.664$$\pm{0.457}(3\sigma)$~&$32.33$~&~$0.874$~\\
\hline $1$+$\eta_a\frac{z}{1+z}$ (Constitution) &$\eta_a$=$-0.793$$\pm{0.436}(3\sigma)$~&$40.46$~&~$1.094$~\\
\hline\hline
\end{tabular}
} \tabcolsep 0pt \caption{\label{Tab1} Fitting results   with the 38
ADDs of galaxy clusters for spherical model and the Union2 set and
Constitution set, and $\chi^2_{\rm min}$ (the minimun $\chi^{2}$),
$\chi^2_{\rm dof}$ ($\chi^{2}_{\rm min}/{\rm dof})$, for
$\eta(z)=1+\eta_1 z$ and $\eta(z)=1+\eta_a\frac{z}{1+z}$,
respectively. The asterisk represents the case with the corrected
luminosity distance interpolated from the nearby SNe Ia.
} \vspace*{5pt}
\end{table}

From Fig 2-3 and Table 1, we have obtained some new results and
insights, which summarized as follows: (1) From comparing to results
of the case with the corrected luminosity distance and the case with
SN Ia set (the Union2 set and the Constitution set) directly, we can
see a shift (significantly with the Constitution set) between the
best fit values and the likelihood contours towards the standard DD
relation ($\eta=1$) with lower $\chi^2_{\rm min}$ for using of the
interpolating method to obtain ${D_L^{\rm corrected}}$. This
situation shows that the using of the interpolating method tend to
avoid the corresponding systematic bias of redshift differences and
make testing results be more compatible with the DD relation. (2)
Compared to results of the case with the Union2 set and the case
with Constitution set, it is shown that that the DD relation of the
one-parameter parameterizations with the Union2 set for the
interpolating method is satisfied within $2\sigma$ CL; while the DD
relation is inconsistent with  the Constitution set for both cases
at $3\sigma$ CL, which shows that the redshift incoincidence of the
observational data pairs can bring the systematic errors and the
systematic errors of the observational data pairs exist
significantly within the Constitution set. For Comparison  with
previous results from the incomplete ADD sample and the Constitution
set \citep{HLR10, Li11}, our analyses with the complete spherical
model sample (38 ADDs) and the Constitution set directly are
consistent with previous results obtained by \citet{HLR10} with the
incomplete spherical model sample (35 ADDs, three points removed by
selection criteria) and the Constitution set, where
$\eta_1=-0.42\pm0.34$, and $\eta_a=-0.66\pm0.50$ at $3\sigma$ CL;
and inconsistent with those obtained by \citet{Li11} with the
incomplete spherical model sample (26 ADDs, 12 points removed by
selection criteria) and the Constitution set, where
$\eta_1=-0.30\pm0.34$ and $\eta_a=-0.46\pm0.51 (3\sigma)$. It
indicates that the choice of selection criteria to remove ADD points
with large bias of redshift differences may play an important role
in testing of the DD relation \citep{CaoL11}, which also means that
the systematic errors brought by redshift incoincidence  exist
significantly within the Constitution set, and the using of the
interpolating method tend to alleviate the corresponding systematic
bias of redshift differences significantly.

\begin{figure*}
\begin{center}
\includegraphics[width=0.39\hsize]{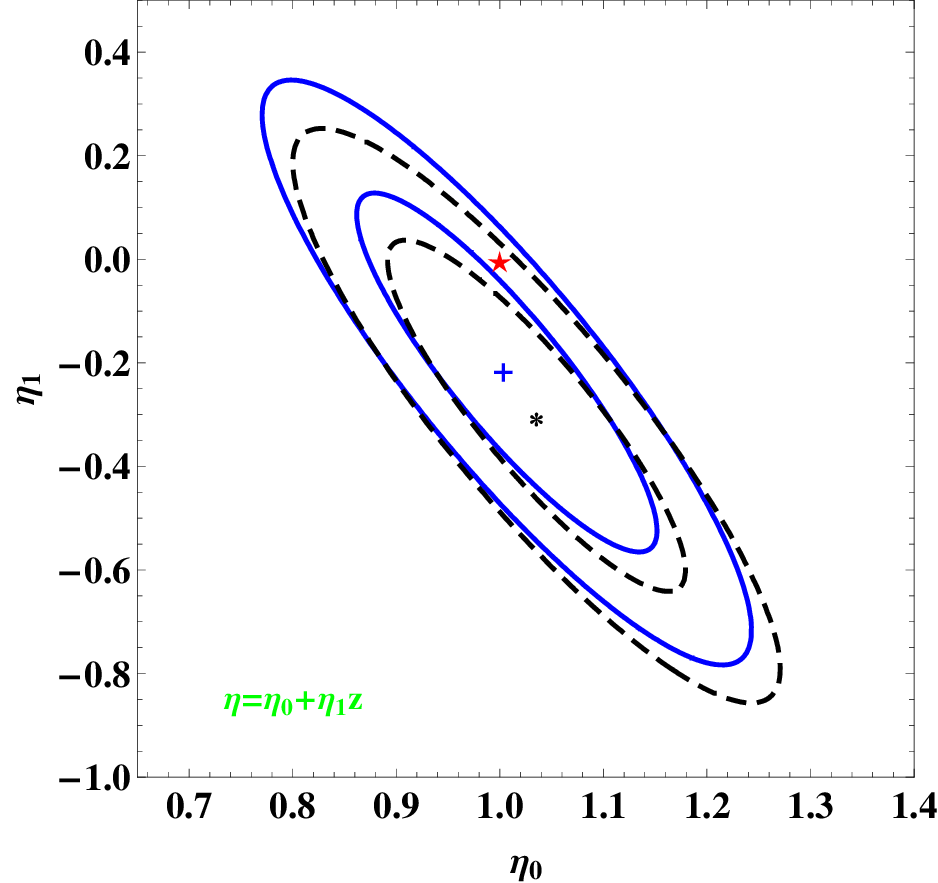}
\includegraphics[width=0.38\hsize]{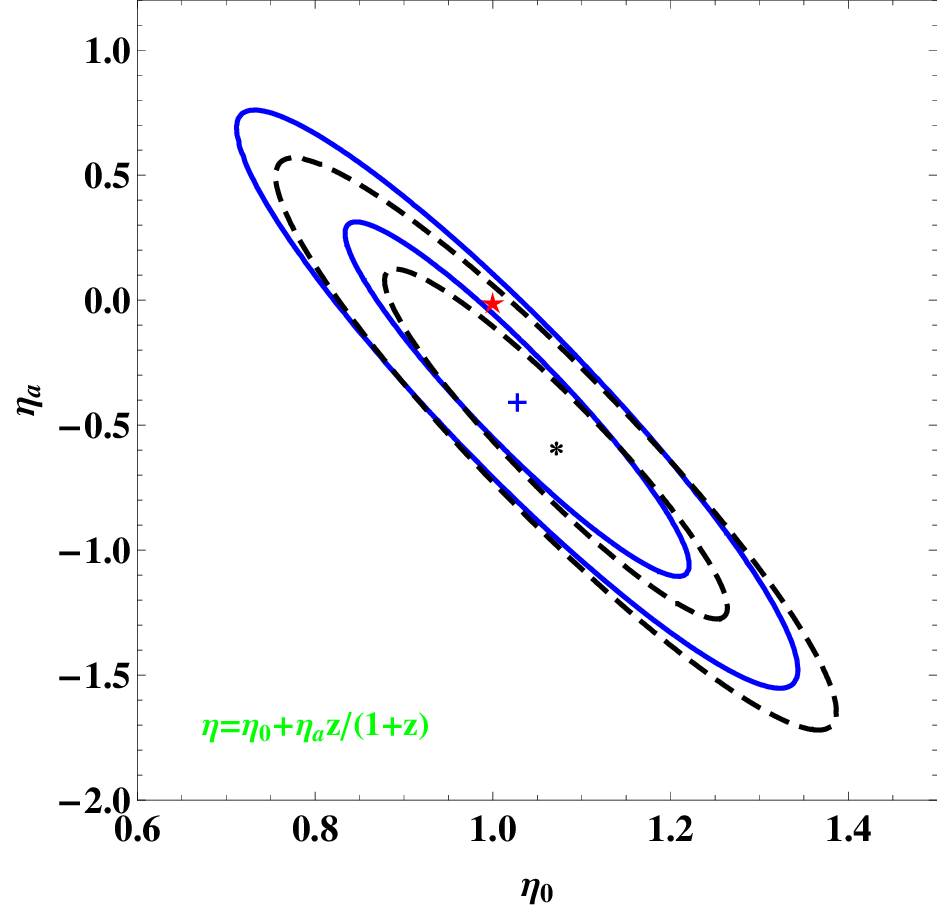}

 \end{center}
\caption{Likelihood contours  with the 38 ADDs of galaxy clusters
and the corrected  luminosity distances of the Union2 set for
two-parameter forms in the $\eta_0-\eta_1$ plane (Left: for
$\eta(z)=\eta_0+\eta_1 z$), and in the $\eta_0-\eta_a$ plane (Right:
for $\eta(z)=\eta_0+\eta_a\frac{z}{1+z}$).  The blue real lines
represent the case with the corrected luminosity distance
interpolated from the nearby SNe Ia, and the black dashed lines
represent the case with the Union2 set directly. The contours
correspond to $1\sigma$ and $2\sigma$ confidence regions, and the
blue crosses and black stars represent the best-fit points and the
red stars represent $\eta(z)=1$.
 \label{fig:2D}}
\end{figure*}

Following \citet{Li11}, we also treat the redshift-independent model
parameter $\eta_{0}$ as a free parameter to examine the DD relation.
Results with two-parameter parameterizations [$\eta (z) = \eta_0 +
\eta_{1}z$ and
$\eta (z) = \eta_0 + \eta_{a}z/(1+z)$] are shown in Figure \ref{fig:2D}. 
With the corrected $D_L$ of the Union2 set, the best-fit values are
$(\eta_0, \eta_1)=(1.007, -0.219)$,  and $(\eta_0, \eta_a)= (1.027,
-0.396)$, respectively. For the case with the Union2 set directly,
the best-fit values are $(\eta_0, \eta_1)=(1.035, -0.302)$, and
$(\eta_0, \eta_a)=(1.071, -0.575)$, respectively. It might be worth
mentioning that marginalizing over $\eta_0$ with two-parameter
parameterizations is similar to adding the $\alpha$ parameter with a
very wide prior.\footnote{We thank the anonymous referee for
pointing this out. The best-fit value of $\eta_0-1$ for the data is
comparable to the width of the prior on $\alpha$ from footnote 1.}
From Figure 4, we can see that a significant shift between the best
fit values and the likelihood contours towards the standard DD
relation ($\eta=1$) for the interpolating method compared to the
testing results of the case with the Union2 set directly.  Our
results suggest that the DD relation for two-parameter
parameterizations are consistent with the observational data
marginally at $1\sigma$ CL, which are more stringent than those
obtained in \citet{Li11}, where the DD relation is  marginally
accommodated at $2\sigma$ CL for two-parameter parameterizations.

\section{Conclusions and Discussions
}\label{sec:Conclusions}

In this work, we preform a new consistent test of the
distance-duality relation [$\eta(z)\equiv D_{L}(1+z)^{-2}/D_{A}$=1]
in a cosmology-independent way. It is obvious that the redshift
differences of observational samples may cause deviation of the DD
relation. Testing results from given pairs of data set with the
corresponding galaxy clusters and SNe Ia at nearby redshift may be
influenced by the particular choice of the selection criteria; the
more stringent selection criteria are used, the more data points
should be removed. In order to avoid any bias of difference of
redshift and ensure the integrity of the ADD samples, we correct the
luminosity distance of a SN Ia to the same redshift of the
corresponding galaxy cluster directly from the nearby SN Ia points.

With the 38 ADD sample from galaxy clusters under an assumption of
spherical model and the corrected luminosity distances of the Union2
set by using the interpolating method  to alleviate the
corresponding systematic bias of redshift differences, fitting
results of the DD relation are $\eta_1$=$-0.232\pm{0.232}$ for
parameterization $\eta(z)$=$1+\eta_a z$, 
and $\eta_a=-0.351\pm{0.368}$  for parameterization
$\eta(z)=1+\eta_a\frac{z}{1+z}$ at $2\sigma$ CL, respectively.  We
also find the DD relation are consistent with the spherical model
and the corrected luminosity distances marginally at $1\sigma$ CL
for the two-parameter forms of parameterization
[$\eta(z)$=$\eta_0+\eta_{a}z$ and
$\eta(z)$=$\eta_0+\eta_{a}z/(1+z)$]. Our results show that there
exists no conceivable evidence for variations in the duality
distance relation when the current SNe Ia (Union2) and the complete
38 sample of galaxy clusters data are confronted, since various
parameterizations of $\eta(z)$ are satisfied within $2\sigma$
CL,\footnote{Note that our results with the 38 ADD sample and the
corrected Union2 set are also consistent with more recent works,
e.g., \cite{Fu11, Meng11, Cardone12}, which considering redshift
bias of observational data pairs to test the DD relation. } which
are more stringent than those obtained in \citet{Li11}, where the DD
relation is only marginally accommodated at $3\sigma$ CL for the
spherical model sample without considering redshift bias. We also
show that the systematic errors brought by redshift incoincidence of
the data pairs exist explicitly within the simulated data by
considering a fiducial cosmological model as well as within the
Constitution set, and the using of the interpolating method tend to
alleviate the corresponding systematic bias significantly.

In the present work, we focus on the 38 galaxy cluster sample under
an assumption of spherical model \citep{Boname06}. When considering
the 25 galaxy clusters under isothermal elliptical model
\citep{DeFilippis05} and the Union2 set, we also find that
$\eta(z)=1$ is well satisfied at $1\sigma$ confidence level for case
with the corrected luminosity distance, which is consistent with
previous works using SN Ia set directly \citep{HLR10, Li11}. This
situation shows that testing results of the DD relation still depend
on the choice of assumptions on the cluster
geometry model significantly. 

Although the excursion of the DD relation  is not significantly
reduced with the corrected sub-sample of Union2 set set and the
complete 38 sample of galaxy clusters data, it should be noted that
the testing results of the DD relation would be improved by
considering
the systematic bias of redshift.

\appendix
\section{The simulated data}
In order to check whether the redshift interpolation indeed removes
the bias of the difference of redshifts between pairs, we test
$\eta(z)$ of parameterizations [$\eta(z)=1+\eta_1z$ and
$\eta(z)=1+\eta_az/(1+z)$] with the simulated data pairs.
In the simulating procedure, we consider the concordance model
($\Omega_{\rm M} = 0.27$, $\Omega_{\rm K} = 0$, $w = -1$, with
Hubble constant $H_0 = 72 {\rm km/s/Mpc}$) as a fiducial
cosmological model to simulate cluster and SNe Ia data.  For the
simulated SNe Ia, we obtain the distance moduli at $\Delta z_{\rm
SN}=0.005$ $(0.14\leq z<0.9)$ from a Gaussian distribution centered
on the theoretical value and with variance $\sigma_\mu=\sigma_{\mu,
\rm obs}(\mu_{\rm sim}/\mu_{\rm obs})$ \citep{Cardone12}. For the
simulated cluster sample, we calculate 38 ADDs from the concordance
model at $z_{\rm cluster}\in[0.142, 0.89]$ from a Gaussian
distribution centered on the theoretical value and with variance
$\sigma_{D_A}=\sigma_{D_A,\rm obs}({D_A}_{\rm sim}/{D_A}_{\rm
obs})$. In the $\chi^2$ fit, we combine the systematic uncertainties
of galaxy clusters and SNe to calculate $\sigma_{\eta_{\rm obs}}$
and produces a reduced $\chi^2$ of unity.

\begin{figure*}
\begin{center}
\includegraphics[width=0.35\hsize]{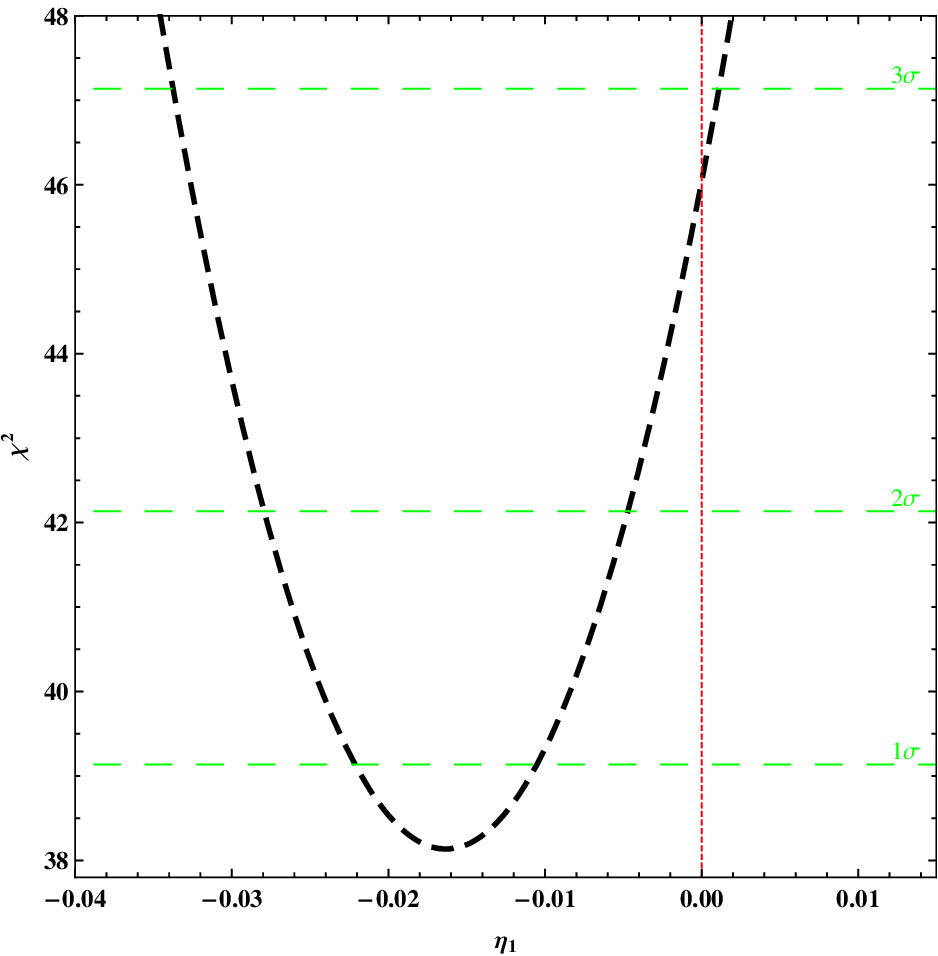}
\includegraphics[width=0.35\hsize]{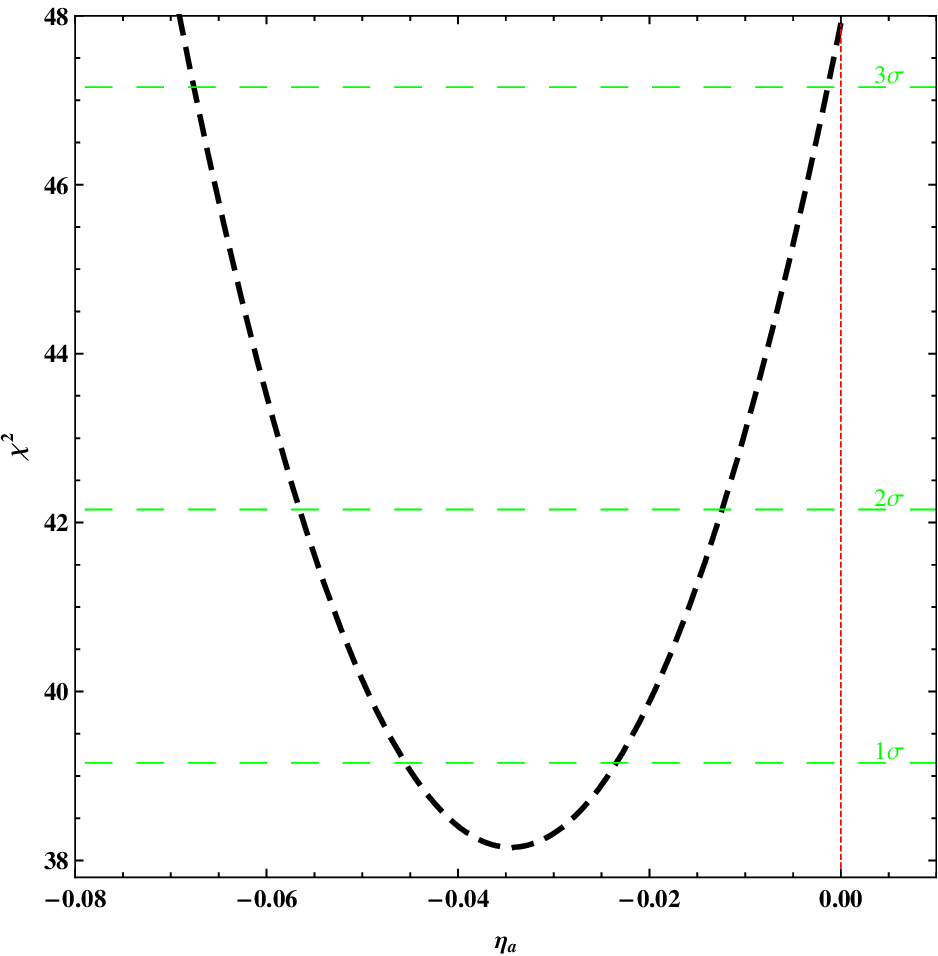}

 \end{center}
\caption{Likelihood contours  with the simulated data pairs  in the
$\eta_1-\Delta\chi^2$ plane (\textit{Left}: for $\eta(z)=1+\eta_1
z$),
 and in the $\eta_a-\Delta\chi^2$ plane (\textit{Right}: for $\eta(z)=1+\eta_a\frac{z}{1+z}$ ).
The black dashed lines represent the case with the simulated  SNe Ia
directly, and the red vertical lines represent $\eta(z)=1$, which
overlap the contours of the case with  the corrected luminosity
distance interpolated from the simulated SNe Ia.
 \label{fig:sim}}
\end{figure*}

We show testing results with the simulated data pairs in Figure
\ref{fig:sim}. For the case with the simulated data pairs directly,
we choose the most close SN point to the redshift of cluster to
build up the pair of galaxy clusters and SNe Ia. The best-fit values
are
$\eta_1=-0.016\pm0.006(1\sigma)\pm0.012(2\sigma)\pm0.017(3\sigma)$
for $\eta(z)=1+\eta_1z$, and
$\eta_a=-0.034\pm0.011(1\sigma)\pm0.022(2\sigma)\pm0.033(3\sigma)$
for $\eta(z)=1+\eta_az/(1+z)$, which shows that $\eta(z)=1$ are
excluded at $2\sigma$  for both cases.  It is indicated that the
difference of redshifts between pairs of galaxy clusters and SNe Ia
could cause deviation in testing the DD relation explicitly.
 For the case with the interpolated SNe data,
we choose the corrected SN point at the same redshift of cluster,
which interpolated from the nearby points, to build up data pairs of
galaxy clusters and SNe Ia. The best-fit values are
$\eta_1=(5.9\pm6.0)\times10^{-5}$ and
$\eta_a=(9.5\pm9.6)\times10^{-5}$ with the $1\sigma$ uncertainties,
which shows that $\eta(z)=1$ is fully satisfied  with the data pairs
interpolated to the same redshifts. Therefore we conclude that the
redshift interpolation method  indeed removes the bias of the
difference of redshifts between data pairs.

\section*{Acknowledgments}
We thank the anonymous referee and Prof. Zhang Tongjie for
constructive suggestions and comments. We also thank  Dr. Hao Wang,
Xiaolong Gong for helpful discussions.
This work was supported by China Postdoctoral Science Foundation Grant No.2012M5102450470237; 
PX. W. acknowledges the National Natural Science Foundation of China
under Grants Nos. 11175093 and 11222545, Zhejiang Provincial Natural
Science Foundation of China under Grants No. R6110518, the FANEDD
under Grant No. 200922; Z-H Z. acknowledges the National Natural
Science Foundation of China under the Distinguished Young Scholar
Grant 10825313, the Ministry of Science and Technology National
Basic Science Program (Project 973) under Grant No.2012CB821804, the
Fundamental Research Funds for the Central Universities and
Scientific Research Foundation of Beijing Normal University.

\end{document}